\documentclass[aps,prb,reprint,superscriptaddress,citeautoscript]{revtex4-1}

\draft 
\usepackage{graphicx}
\usepackage{mathtools}
\usepackage{color,soul}
\usepackage[framemethod=tikz]{mdframed}

\begin{document}


\title{Enhanced quasiparticle lifetime in a superconductor by selective blocking of recombination phonons with a phononic crystal} 



\author{K. Rostem}
\email[]{karwan.rostem@nasa.gov}
\affiliation{NASA Goddard Space Flight Center, 8800 Greenbelt Road, Greenbelt, MD 20771, USA}

\author{P. J. de Visser}
\affiliation{SRON, Netherlands Institute for Space Research, Sorbonnelaan 2, 3584 CA, Utrecht, The Netherlands}

\author{E. J. Wollack}
\affiliation{NASA Goddard Space Flight Center, 8800 Greenbelt Road, Greenbelt, MD 20771, USA}


\date{\today}

\begin{abstract}

When quasiparticles in a BCS superconductor recombine into Cooper pairs, phonons are emitted within a narrow band of energies above the pairing energy at 2$\Delta$. We show that a phonon bandgap restricting the escape of recombination phonons from a superconductor can increase the quasiparticle recombination lifetime by more than an order of magnitude. A phonon bandgap can be realized and matched to the recombination energy with a phononic crystal, a periodically-patterned dielectric membrane. We discuss in detail the non-equilibrium quasiparticle and phonon distributions that arise in a superconductor due to a phonon bandgap and a pair-breaking photon signal. Although intrinsically a non-equilibrium effect, the lifetime enhancement in the small-signal regime is remarkably similar to an estimate from an equilibrium formulation. The equilibrium estimate closely follows $\exp(\Omega_{bg}/k_BT_b)$, where $\Omega_{bg}$ is the phonon bandgap energy bandwidth above 2$\Delta$, and $T_b$ is the phonon bath temperature of the coupled electron-phonon system. We discuss the impact of a phononic bandgap on the performance of a superconducting circuit element, and propose a microwave resonator to measure the enhancement in the quasiparticle lifetime. 

\end{abstract}


\maketitle 

\section{Introduction}

Nano-patterned dielectric membranes have enabled devices in which the coupling of phonons, photons, and electrical signals are managed with exquisite sensitivity~\cite{alegre,QIprocessing,Chan-AcousticShield}. The elastic properties of the patterned dielectric, which is generally referred to as a phononic crystal (PnC), can be significantly different from the bulk material. In particular, a PnC infinite in extent can possess a complete phonon bandgap, a property that has been studied extensively to control phonon interactions~\cite{30percent,alegre,amir:2010,phononic:transistor,Yu:phononicShield} and reduce heat transport at room temperature and in cryogenic devices~\cite{zen,rostem-aperiodic,anufriev-meas}. A PnC bandgap has been demonstrated to reduce the energy loss of a mechanical resonator to a phonon bath~\cite{Chan-AcousticShield}.

\begin{figure*}[htbp]
\begin{center}
\includegraphics[width=16cm]{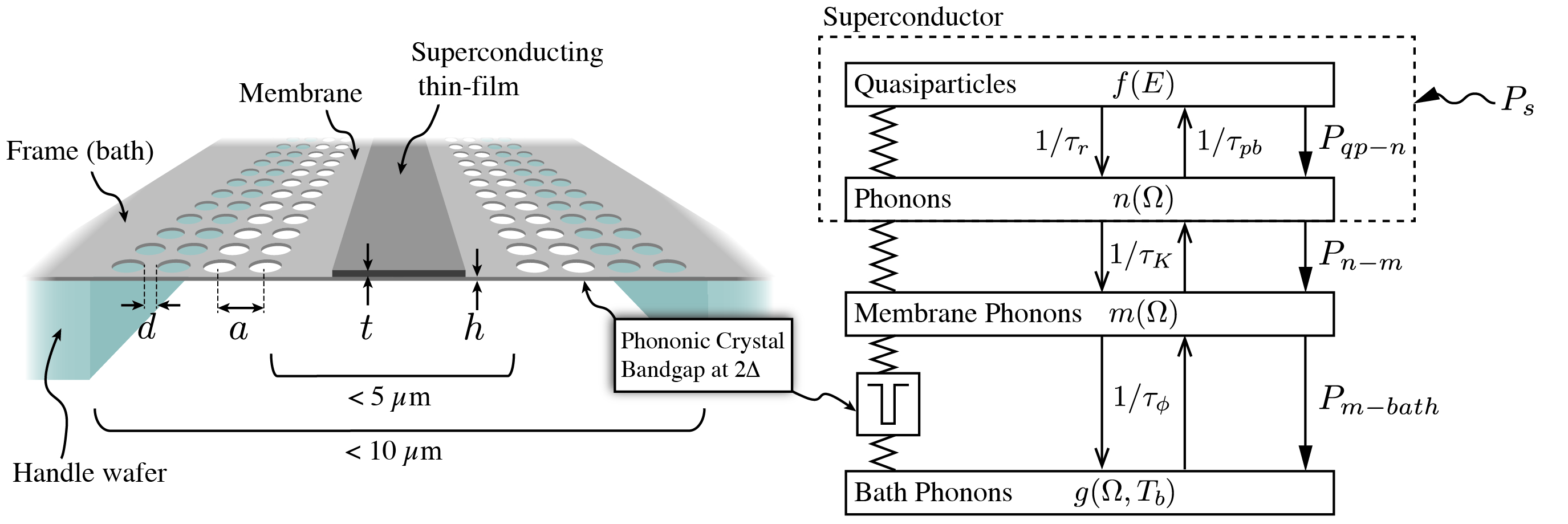}
\caption{Schematic of a phononic-isolated superconducting film and the corresponding electro-thermal model. The phononic crystal pattern of circular holes etched in a square tiling is shown for reference. The phononic crystal unit cell geometry and tiling can be modified to suit the application. Typical dimensions are shown for a superconducting thin-film realized as the inductor element of a superconducting microwave resonator. $P_s$ is a pair-breaking photon signal power. In the simulations presented here, the bath temperature, $T_b$, is set to $0.1\,T_c$, where $T_c$ is the critical temperature of the superconducting film. Other parameters are described in the text. }
\label{fig:KID-model}
\end{center}
\end{figure*}

Electron-phonon coupling is intrinsic to superconductivity. When two quasiparticles recombine to form a Cooper pair in a superconductor, a phonon is emitted at the pairing energy, $2\Delta$. A recombination phonon can either remain in the superconducting film and break another pair, or escape to the dielectric substrate (see Fig.~\ref{fig:KID-model}). The phonon pair-breaking and escape rates, together with the electron-phonon coupling strength, determine the quasiparticle lifetime, $\tau_{qp}$, in the superconductor~\cite{chang_NE,wilson}. A PnC bandgap centered at 2$\Delta$ should substantially reduce the escape rate of recombination phonons from the superconductor, and increase $\tau_{qp}$. 

The quasiparticle recombination lifetime plays an important role in noise processes of pair-breaking superconducting devices. In a kinetic inductance detector (KID~\cite{day,jonas}), the fundamental limiting noise in steady-state is from generation and recombination (GR) of quasiparticles~\cite{deVisser-Ncomms}. The noise-equivalent power (noise power detected in a 1 Hz bandwidth) due to GR fluctuations is proportional to $\sqrt{1/\tau_{qp}}$. In a quantum capacitance detector, an increase in $\tau_{qp}$ would reduce telegraph noise associated with quasiparticle tunneling through a Josephson junction~\cite{QCD-theory,QCD-2018}. Independent of the fluctuation mechanism limiting the device noise, a reduction in the recombination phonon escape rate would lead to an increase in the responsivity of a superconducting thin-film to pair-breaking photons. By symmetry of the phonon transport across a PnC (a passive linear system), a superconducting circuit can potentially be an effective shield against quasiparticle poisoning induced by pair-breaking phonon injection into a superconductor~\cite{QUBIT-poisening}.

A superconducting resonator fabricated on a thin ($<$1 $\mu$m) self-supporting dielectric membrane has been suggested for increasing $\tau_{qp}$~\cite{vercruyssen, fyhrie}. The membrane dimension is significantly larger than the phonon mean-free-path ($<\,$100 $\mu$m)~\cite{klitsner,holmes,hoevers,zen}. Thus, phonons in the membrane are weakly coupled to the thermal bath, and the phonon lifetime is increased at all energies. Additionally, the quasiparticle and phonon occupation in the superconducting film approach a thermal distribution with an effective temperature corresponding to the signal power, and $\tau_{qp}$ approaches the thermal response time $\tau_{th}=C/G$, where $C$ is the total heat capacity from quasiparticles and phonons, and $G$ is the thermal conductance to the bath~\cite{vercruyssen}. Hence, such a membrane-isolated superconducting resonator functions as a bolometer~\cite{lindeman}. 

In this paper, we examine the electron-phonon system of a superconducting thin-film surrounded by a PnC, see Fig.~\ref{fig:KID-model}. When the phonon bandgap of the PnC is matched to the pair-breaking frequency of the superconductor at $\sim\,$2$\Delta$, the quasiparticle lifetime becomes a strong function of the bandgap properties. The PnC only increases the lifetime of membrane phonons with energies within the bandgap. Above and below the bandgap energies, the phonon transmission to the bath often rapidly approaches unity~\cite{hou,30percent}. By design, the length of the membrane and PnC combined is less than the phonon mean-free-path~\cite{klitsner,holmes,hoevers,zen}, and consequently thermal phonons ($<$10 GHz) in the membrane supporting the superconducting film remain tightly coupled to the bath. The quasiparticle and phonon distributions in the system will be out of equilibrium since the thermal phonon escape rate is comparable to or greater than the pair-breaking rate.

We solve the coupled kinetic equations of the electron-phonon system in Fig.~\ref{fig:KID-model} to estimate the quasiparticle lifetime enhancement due to a PnC bandgap. The solution takes into account electron-phonon scattering in the superconductor, pair-breaking due to phonons and incident signal photons, and phonon transport as modified by a PnC bandgap. 

\section{\label{sec:noneq} Kinetic equations and non-equilibrium distributions}

The quasiparticle and phonon distributions of the system in Fig.~\ref{fig:KID-model} will in general be out of equilibrium when the system is driven by an external source. For example, in the presence of a monochromatic pair-breaking photon signal of energy $h\nu_s$, the quasiparticle distribution peaks at $h\nu_s-\Delta$ and $h\nu_s+\Delta$, which correspond to pair-breaking and quasiparticle photon absorption~\cite{chang_KE,goldie,guruswamy_eta}, respectively. The phonon distributions have at least one peak at 2$\Delta$ due to quasiparticle recombination. The non-equilibrium particle distributions can be determined by solving the Boltzmann kinetic equations that describe the rate of change of the distributions under phonon emission and absorption by a quasiparticle, pair-breaking by photons and phonons~\cite{chang_KE}. 

\subsection{Rate Equations}
The rate equations for the quasiparticle and phonon energy distributions in Fig.~\ref{fig:KID-model} are
\begin{eqnarray}
\dot{f}(E) &=& I_{qp}(E,\Omega_s) - \frac{1}{\tau_0(k_BT_c)^3} (I_1 + I_2 + I_3), \label{eqn:f_KE} \\
\dot{n}(\Omega) &=& -\frac{2}{\pi\tau^\phi_0\Delta(0)}  (J_1 + J_2/2) - \frac{n(\Omega)-m(\Omega)}{\tau_{K}}, \\
\dot{m}(\Omega) &=& -\frac{m(\Omega)-n(\Omega)}{\tau_{K}} - \frac{m(\Omega)-g(\Omega,T_b)}{\tilde{\tau}_{\phi}(\Omega)} \label{eqn:m_KE}, 
\end{eqnarray}
where $I_k$ and $J_k$ describe electron-phonon scattering and pair-breaking events (see Appendix for details). $\tau_0$ and $\tau^\phi_0$ are intrinsic energy-independent quasiparticle and phonon lifetimes in the superconductor~\cite{kaplan}. $I_{qp}$ in Eq.~\ref{eqn:f_KE} describes the quasiparticle injection rate from photon absorption. The absorbed photon power, $P_{s}$, terminates in a phonon thermal bath. Power flow through the system defines the energy conservation equations in steady-state, which are solved simultaneously with Eq.~\ref{eqn:f_KE}~to~\ref{eqn:m_KE} to find a self-consistent solution for $f(E)$, $n(\Omega)$, and $m(\Omega)$. We utilize the iterative technique described in Ref.~\onlinecite{goldie} to solve the kinetic and power flow equations for small and large-signal pair-breaking photon loading conditions. Only pair-breaking photons are considered here, though adding a term for microwave photon injection is straightforward~\cite{guruswamy_eta}. It is well known that microwave readout photons in a superconducting thin-film resonator can lead to pair-breaking, which arises from repeated photon absorption by quasiparticles~\cite{goldie}. We neglect this effect here to purely assess the physics of recombination in the presence of a phonon bandgap. 

\subsection{\label{sec:bandgap}PnC bandgap model}

The effect of a phonon bandgap is introduced in Eq.~\ref{eqn:m_KE} through the membrane phonon lifetime, $\tilde{\tau}_\phi(\Omega) = \tau_0^\phi\,r(\Omega)$, where 
\begin{equation}
r(\Omega) =
\begin{cases}
r_{bg} & \Omega = \text{[$\Omega_{bg}$\,-\,$B_{bg}$/2, $\Omega_{bg}$\,+\,$B_{bg}$/2]}, \\
1 & \text{otherwise}. \\
\end{cases}
\label{eqn:rbg}
\end{equation}
$\tau_0^\phi$ is an intrinsic phonon lifetime in the superconductor related to the electron-phonon coupling strength~\cite{kaplan}. $\Omega_{bg}=h\nu_{bg}$, $B_{bg}$, and $r_{bg}$ are the center energy, bandwidth, and rejection level of the PnC bandgap. As a rule of thumb for the design of a PnC, $\Omega_{bg}$ is approximately linearly dependent on the lattice constant, $a$, of the PnC (see Fig.~\ref{fig:KID-model}). $\nu_{bg}a/v_s = q$, where $v_s \sim 5000$ ms$^{-1}$ is the bulk shear mode sound velocity of the PnC parent material (e.g. single-crystal silicon), and $q$ is a constant that depends on the details of the unit cell geometry. For circular holes etched in a square tiling, $q\sim0.6$ when the hole radius $r$ is 0.45$a$. The minimum feature size, $d=a-2r$, sets the highest phonon bandgap center frequency that can be achieved in practice. Tens of nanometer features can be realized with electron-beam lithography, suggesting a reasonable practical bandgap upper limit is $\nu_{bg}\sim20$ GHz~\cite{anufriev-theory}. Bandgaps at higher frequencies, $\sim\,$30 GHz, can be generated with more complex geometries~\cite{Tuomas}. 

In using the relation $\tilde{\tau}_\phi(\Omega) = \tau_0^\phi\,r(\Omega)$, we assume the phonon escape time to the bath at energies above and below the bandgap of the PnC is on the order of the escape time from the superconductor to the bath ($\tau_K\simeq\tau_0^\phi$)~\cite{goldie}. This model is sufficient to describe the effect of a phonon bandgap on the quasiparticle recombination time, which is captured as a normalized parameter that is independent of the values of $\tau_0$, $\tau_0^\phi$, $\tau_{K}$, and $\tilde{\tau}_\phi(\Omega)$.

\subsection{Superconductor model and material parameters}
We use the well known materials properties and low-temperature superconducting parameters of Al~\cite{kaplan} to model the effect of a phonon bandgap on the quasiparticle lifetime, $\tau_{qp}$ (see Appendix and Table~\ref{tbl:summary} therein). There are two inherent assumptions in using Al: (1) The superconductor quasiparticle density of states is described by Bardeen-Cooper-Schrieffer (BCS~\cite{BCS}) theory, $\rho_{qp}(E) = \Re[E/\sqrt{E^2-\Delta^2}]$, which ensures the particle dynamics are dominated by phonons with energies near $2\Delta$, and (2) the electron-phonon interaction function $\alpha^2(\Omega)\rho_\phi(\Omega)$ is approximated by $b\,\Omega^2$, where $\rho_\phi(\Omega)$ is the phonon density of states and $b$ is a material-dependent constant~\cite{kaplan}. 

Matching the PnC bandgap to the energy gap of the superconductor, $\Omega_{bg} \sim 2\Delta$, and using the BCS relation $2\Delta = 3.53k_BT_c$, suggests materials with $T_c < 0.3$ K are suitable for demonstrations of a PnC-isolated superconducting device. In the model described here, $T_c$ merely scales the energy gap of the superconductor.

\begin{figure}[t]
\begin{center}
\includegraphics[width=7.5cm]{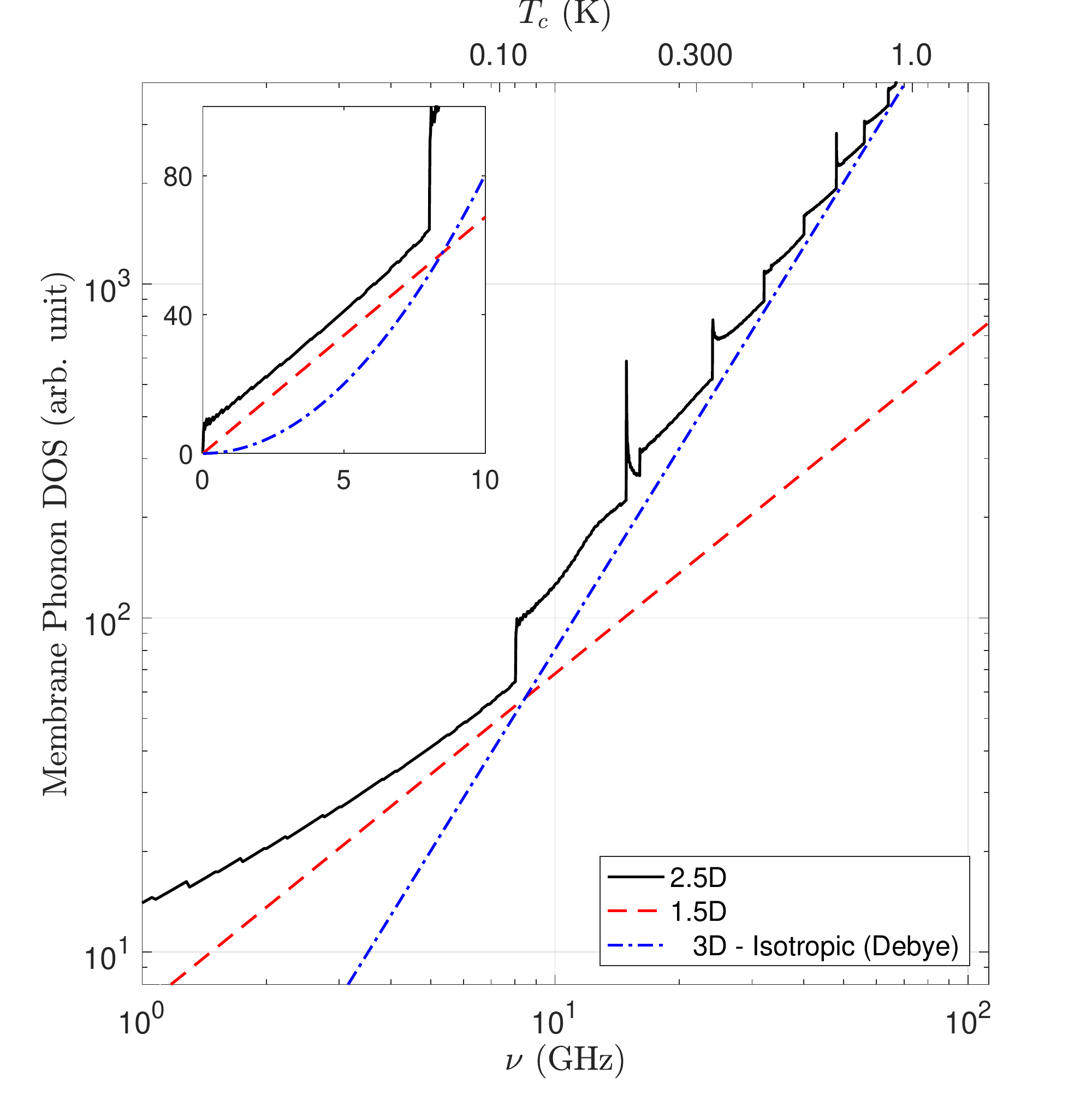}
\caption{Phonon density of states of a uniform elastic membrane (solid line). The phonon modes are the symmetric and antisymmetric Lamb modes, and a shear horizontal mode. D represents an approximate dimensionality for the phonon modes of a uniform plate. $\rho_\phi^{2.5\rm D} = \rho_\phi^{2\rm D}/h$, where $h$ is the membrane thickness. $\rho_\phi^{2.5\rm D}$ approaches the Debye approximation in the 3D limit ($\rho_\phi^{3 \rm D} = 12\pi\nu^2/v_a^3$, where $v_a = \sqrt[3]{3/[2/v_s^3 + 1/v_l^3]}$ is an average sound velocity, and $v_s$ and $v_l$ are the bulk shear and longitudinal mode velocities; dot-dashed line). $\rho_\phi^{2.5\rm D}$ is approximately linear with frequency at the long-wavelength limit ($\rho_\phi^{1.5\rm D} = \rho_\phi^{1\rm D}/h = 6\pi\nu/v_a^2/h$, where $v_a = \sqrt{3/[2/v_s^2 + 1/v_l^2]}$ in this case; dashed line). }
\label{fig:phonon_dos}
\end{center}
\end{figure}

It is instructive to consider the validity of the Debye approximation $\alpha^2(\Omega)\rho_\phi(\Omega)$ as a function of $T_c$. Figure~\ref{fig:phonon_dos} shows the phonon density of states of a 200 nm elastic membrane. For simplicity, we treat the superconducting metal and dielectric membrane as one elastic structure, and calculate the density of states due to Lamb modes and the shear horizontal modes of a thin elastic plate. At low frequencies, only three acoustic phonon modes exist, and $\rho_\phi(\Omega)$ is approximately linear with frequency~\cite{zen}. Above a critical frequency that is characteristic of the bulk shear sound speed-to-membrane thickness, $\rho_\phi(\Omega)$ approaches the Debye model for a bulk elastic medium. Thus, the approximation $b\,\Omega^2$ is reasonable for superconducting metals with $T_c > 0.3$ K (or pair-breaking energy $> 20$ GHz), assuming a typical shear sound speed ($\sim$5000 km/s) and membrane thickness ($\sim$100 nm). For $T_c<0.3$ K, the number and dispersive nature of the phonon modes in the superconducting film may be needed to accurately evaluate $\alpha^2(\Omega)\rho_\phi(\Omega)$. 

\subsection{Steady-State Solutions}

\begin{figure}[t]
\includegraphics[width=7.5cm]{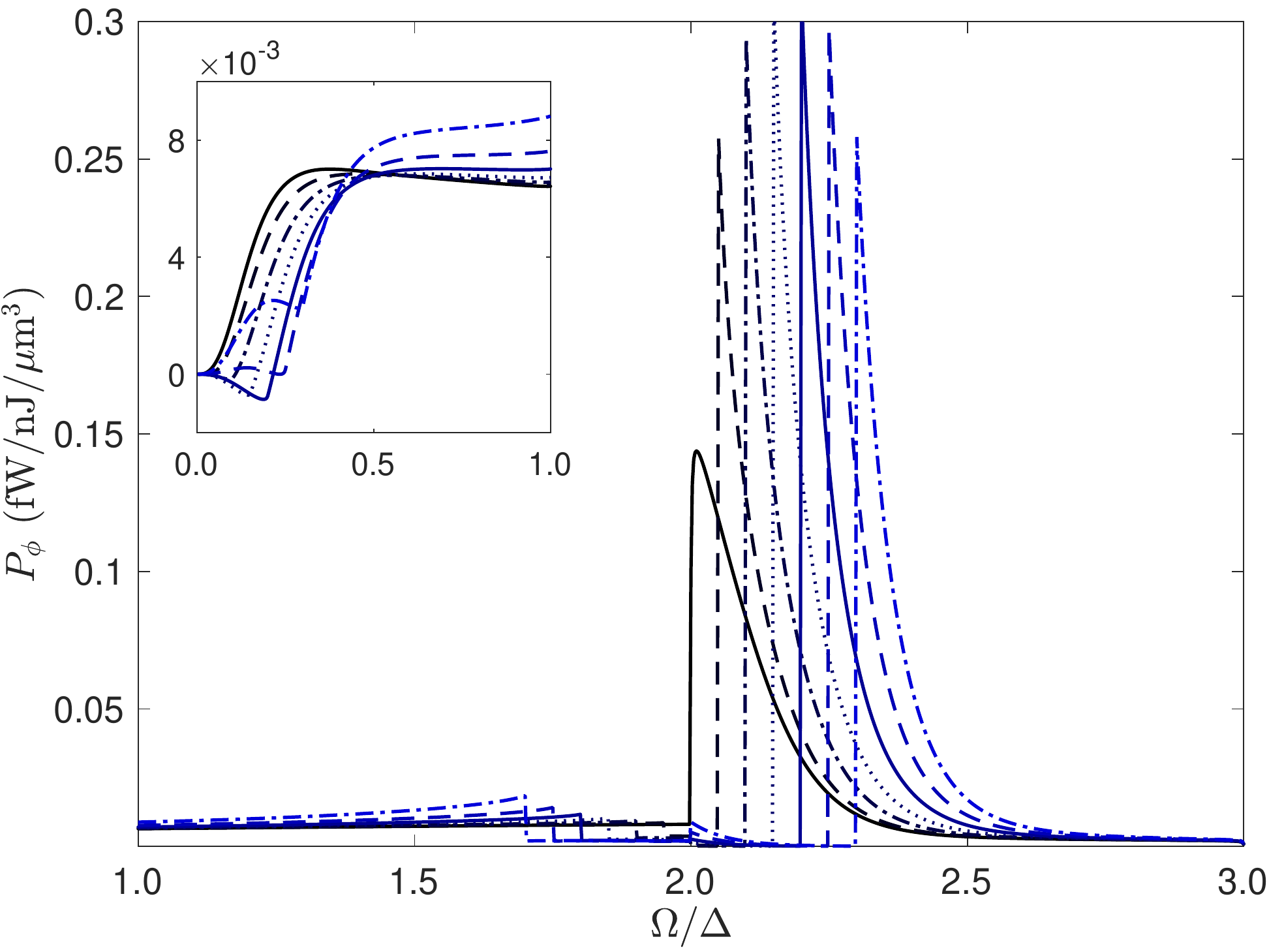}
\includegraphics[width=7.5cm]{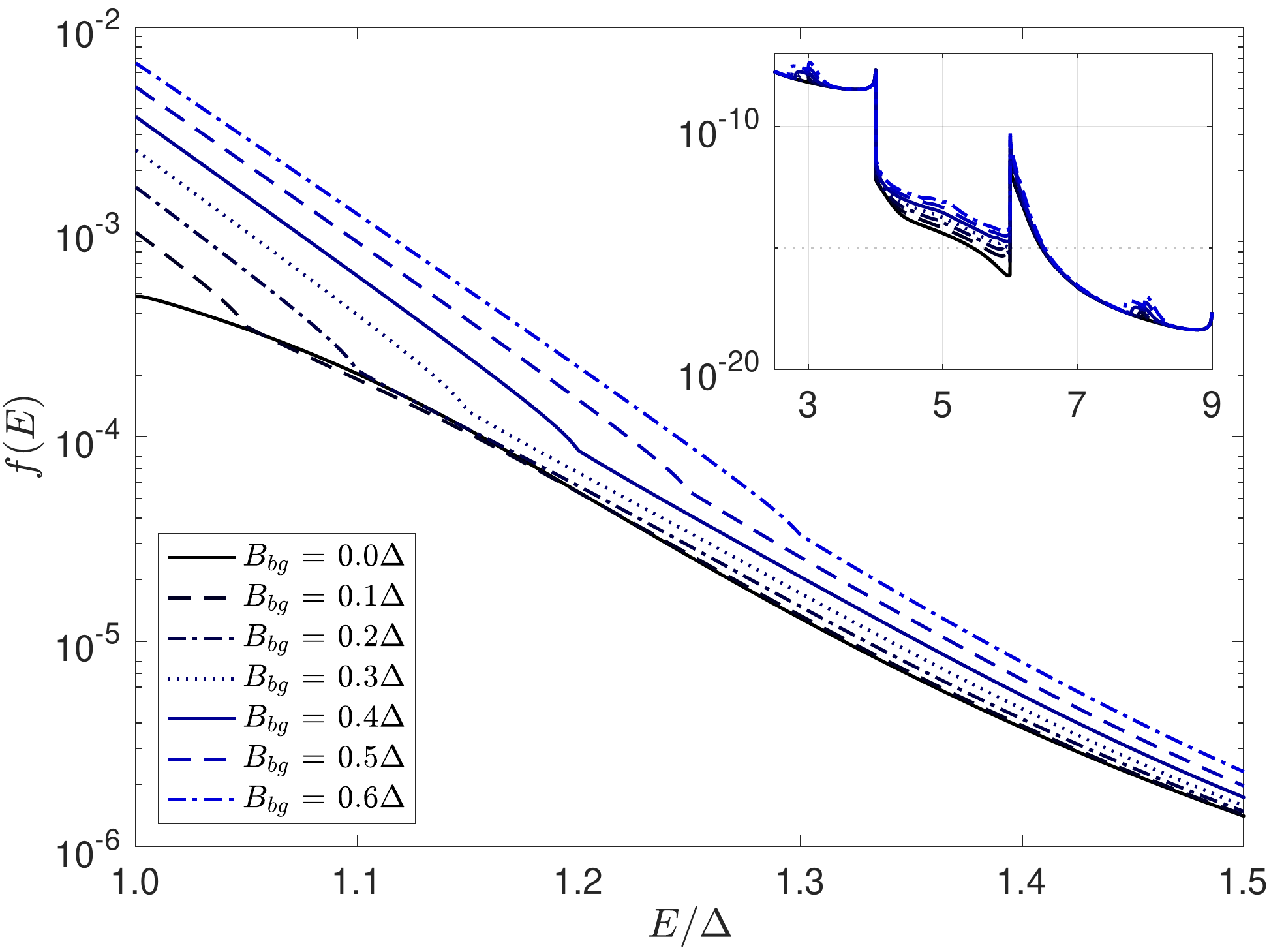}
\caption{\label{fig:dist}(Left) Power spectrum of the phonon power in the electron-phonon system of Fig.~\ref{fig:KID-model} (integrand of Eq.~\ref{eqn:Ps-qp} to \ref{eqn:Pm-bath}). The phonon bandgap rejection, $r_{bg}$, is $10^4$. The bandgap center is at 2$\Delta$. The signal frequency is 5$\Delta$, and signal power is 1 fW/$\mu$m$^3$. The inset shows the spectrum at low frequencies, where the power spectrum can be negative as described in the text. (Right) The corresponding quasiparticle distribution. Since the bandgap is centered at 2$\Delta$, $f$ is only significantly enhanced up to $B_{bg}/2$. The inset shows the increase in the occupation at 3 and 8 $\Delta$, which is due to up-scattering from recombination phonons. }
\end{figure}

The non-equilibrium quasiparticle distribution and power spectrum of the phonon flux are shown in Fig.~\ref{fig:dist}. In this model, $\Omega_{bg} = 2\Delta$, $B_{bg} = 0.3\,\Omega_{bg}$, and $r_{bg} = 10^4$, properties that can be achieved readily with a PnC. As expected, the phonon power is reduced within the bandgap bandwith. 

The signal power, injected at 5$\Delta$, is redistributed through electron-phonon interaction, predominantly producing quasiparticles with energies close to $\Delta$. For pairs of quasiparticles with a total energy less than 2$\Delta$+$B_{bg}/2$, recombination is not effective since a phonon at these energies is more likely to remain in the superconductor than to propagate to the bath. Recombination therefore only takes place effectively for pairs of quasiparticles with a total energy larger than 2$\Delta$+$B_{bg}/2$. Consequently the peak phonon flow is shifted to the upper edge of the phononic bandgap. As a secondary effect, the preference for recombination of higher energy quasiparticles enhances the chance for quasiparticles very close to $\Delta$ to absorb low energy phonons, which have a high occupation. The net phonon flow to the bath at energies $\Omega \simeq [0,B_{bg}/2]$ is reduced, and can even be negative (see inset to Fig.~\ref{fig:dist}), meaning a net phonon in-flow occurs from the bath to the superconductor at these low energies. Additionally, the increasing number of 2$\Delta$ phonons in the superconducting film up-scatter quasiparticles to higher energies. This effect is most noticeable where the quasiparticle density of states peaks, as 1 and 6 $\Delta$ quasiparticles are up-scattered to 3 and 8 $\Delta$, respectively. The spectral width of the feature centered at 3 and 8 $\Delta$ is equal to $B_{bg}$. For the highest signal power loading explored here, the quasiparticle distribution leads to a maximum of 4\% change in the energy gap from the zero-temperature value. This perturbative change in $\Delta$ as a function of signal power is neglected.

For photon energies significantly above 2$\Delta$, the response of a superconducting film is expected to be a stronger function of signal power than signal energy. The effect of the latter is largely captured in the optical pair-breaking efficiency factor, $\eta_{pb}$~\cite{guruswamy_eta}. In the models presented here, the signal photon energy is kept constant and equal to 5$\Delta$. On the other hand, the signal power, $P_s$, directly affects the number of quasiparticles in the superconductor, and the number of 2$\Delta$ phonons in the system. Thus, we vary the signal power, $P_s$, and parameters associated with the PnC bandgap, which include the bandgap center energy, $\Omega_{bg} = h\nu_{bg}$, bandwidth $B_{bg}$, and rejection level, $r_{bg}$ (see Eq.~\ref{eqn:rbg}).

\section{\label{sec:chi} Quasiparticle lifetime enhancement}
\subsection{Non-equilibrium}

Once the distribution functions are known, the enhancement of the phonon lifetime in the membrane due to the phonon bandgap can be calculated from a set of modified Rothwarf-Taylor (RT) equations~\cite{RT}. This formulation essentially captures an energy-dependent phonon lifetime function given by Eq.~\ref{eqn:rbg} as a single effective parameter of the electron-phonon system in Fig.~\ref{fig:KID-model}. The enhancement in the quasiparticle lifetime due to the bandgap can be consistently calculated within this framework. For the system in Fig.~\ref{fig:KID-model}, The RT equations are
\begin{eqnarray}
\dot{N_{qp}} &=& \Gamma_s - {\Gamma_r(N_{qp})N_{qp}} + 2\Gamma_{pb}N_{\omega} \label{eqn:RT1} \\
\dot{N_{\omega}} &=& {\Gamma_r(N_{qp})N_{qp}/2} - \Gamma_{pb}N_{\omega} - \Gamma_{K}(N_{\omega} - N_{\omega}^m) \\
\dot{N_{\omega}^m} &=& \Gamma_{K}(N_{\omega} - N_{\omega}^m) - \Gamma_{\phi}(N_{\omega}^m - N_{\omega}^B) \label{eqn:RT1_1}
\end{eqnarray}
where $N_{qp}=4N(0)\int_\Delta^\infty \rho_{qp}(E,\Delta)\,f(E)dE$ is the number of quasiparticles in the superconductor, and $N(0)$ is the single-spin density of states at the Fermi energy. $N_{\omega}=\int_{2\Delta}^{\infty} \rho_{\phi}(\Omega)n(\Omega)d\Omega$ is the number of $\ge$$2\Delta$ phonons in the superconducting film. Similarly, $N_{\omega}^m$ and $N_{\omega}^B$ are $\ge$$2\Delta$ phonons in the membrane and thermal bath, respectively. $\Gamma_s= \eta_{pb} P_s/\Delta$, is the quasiparticle injection rate due to pair-breaking signal photons, where $P_s$ is the signal photon power, and $\eta_{pb}$ is the photon pair-breaking efficiency~\cite{guruswamy_eta}.

$\Gamma_r(N_{qp}) = 1/\tau_{r} = 2RN_{qp}$ is the quasiparticle recombination rate, where $R$ is a distribution-averaged recombination rate as derived in Ref.~\onlinecite{chang_KE}. Similarly, $\Gamma_{pb}$ is a distribution-averaged pair-breaking rate~\cite{chang_KE}. $\Gamma_K$ is the phonon escape rate from the superconducting film to the membrane. $\Gamma_r$, $\Gamma_{pb}$, and $\Gamma_K$ represent processes that occur, on average, near the energy scale of $\Delta$ for recombination and 2$\Delta$ for phonons due to the BCS quasiparticle density of states~\cite{BCS}. $\Gamma_{\phi}=1/\tau_{\phi}$ is an energy-independent phonon escape rate from the membrane to the bath, and quantifies the effect of a phonon bandgap. 

In steady state, $\dot{N_i}=0$, and assuming $N_\omega^m$\,$\gg$\,$N_\omega^B$, Eq.~\ref{eqn:RT1}~to~\ref{eqn:RT1_1} can be solved for $N_{qp}$ to give
\begin{eqnarray}
N_{qp} &=& \Gamma_s\,\tau_{r}\,\left(1+\frac{\tau_{K}}{\tau_{pb}} + \frac{\tau_{\phi}}{\tau_{pb}}\right), \label{eqn:Nqp_solved}\\
	    &=& \Gamma_s\,\tau_{r}\,F_\phi . \nonumber 
\end{eqnarray}
The second and third terms in the brackets arise due to phonon trapping in the superconductor and the membrane, respectively. Thus, the recombination lifetime is enhanced when one or both of the phonon lifetimes in the system significantly exceed the pair-breaking lifetime. When the membrane phonons are strongly coupled to the bath, $\tau_{\phi} \ll \tau_{pb}$, and the recombination time enhancement due to boundary resistance is recovered, $F_\phi = (1+{\tau_{K}}/{\tau_{pb}})$ ~\cite{wilson,chang_NE}. 

To characterize the effects of a phonon bandgap on the recombination lifetime, we define $F_\phi = 1+({\tau_{K}} + {\tau_{\phi}})/{\tau_{pb}}$, and an enhancement $\chi \equiv {F^*_{\phi}}/{F_{\phi}}$, where $F^*_\phi$ is $F_\phi$ evaluated with a phonon bandgap present in the system. $\chi$ defines the enhancement for a constant quasiparticle injection rate, and describes in a general sense (under equilibrium and non-equilibrium conditions) the effects of an energy-dependent phonon lifetime in the electron-phonon system of a superconductor on the quasiparticle lifetime. Substituting $\tau_{r} = 1/2RN_{qp}$ into Eq.~\ref{eqn:RT1_1} and solving for $F_\phi$ gives an equivalent expression for the enhancement parameter, $\chi = {R^*N_{qp}^{*2}}/{RN_{qp}^{2}}$. 

We note that the measurable small-signal quasiparticle lifetime is $\tau_{qp}^0=\tau_r(N_{qp}^0)F_\phi/2$, where the factor of 1/2 appears due to the linearization of the quadratic term in $N_{qp}$ in Eq.~\ref{eqn:RT1}. $N_{qp}^0$ is the equilibrium quasiparticle number density in the absence of the small pair-breaking signal. 

Equation~\ref{eqn:Nqp_solved} can be solved to give $\eta_{pb} = 2R\Delta N_{qp}^2/F_\phi P_s$, which is intrinsically a distribution-averaged parameter. The change in $\eta_{pb}$, $R$ and $\Gamma_{pb}$ as a function of bandgap and pair-breaking signal photon power is small ($<$20\%) relative to the enhancement in the quasiparticle lifetime, which is at least an order of magnitude for reasonable phonon bandgap properties (see Sec.~\ref{sec:results} and Fig.~\ref{fig:chi}).

\subsection{Equilibrium approximation}
It is instructive to first highlight an approximation that is based on equilibrium distributions. The recombination time for a single quasiparticle at energy $E$ was derived by Kaplan et al. \cite{kaplan}, averaging over all possible phonon energies. In this formulation, it is possible to implement the effect of a phonon bandgap by restricting the energies over which recombination phonons can be released. The equation for the recombination time is then given by:
\begin{eqnarray}
\frac{1}{\tau_{r}(E)} &=& \frac{1}{\tau_0(k_BT_c)^3} \int_x^{\infty} \frac{\Omega^2 \rho_{qp}(\Omega-E)}{[1-f(E,T_b)]}\left(1+\frac{\Delta^2}{E(\Omega-E)}\right) \nonumber \\
			     &\times& [n(\Omega,T_b)+1]f(\Omega-E,T_b) d\Omega,\label{eqn:taurKaplanE}
\end{eqnarray}
where in particular the lower limit $x$ of the integral is changed from $E+\Delta$ to $\max(E+\Delta , 2\Delta+B_{bg})$, which effectively forbids recombination of quasiparticles by emission of a phonon that has its energy within the phononic bandgap. This equation is often evaluated using $E=\Delta$, which is only a good approximation for a thermal electron and phonon distribution. To account for a phonon bandgap, which changes the recombination time over a broad energy range, we calculate a recombination time for an ensemble of quasiparticles averaged over all energies,
\begin{equation}
\left<\tau_{r}\right> = \frac{\int_{0}^{\infty} \tau_{r}(E) f(E,T_b) \rho_{qp}(E) dE} {\int_{0}^{\infty} f(E,T_b) \rho_{qp}(E) dE} ,\label{eqn:taurKaplanaverage}
\end{equation}
In this equilibrium approximation, the quasiparticle lifetime enhancement factor is $\chi_r = \left<\tau^*_{r}\right>/\left<\tau_{r}\right>$, where the Asterix again defines the recombination lifetime in the presence of a phonon bandgap in the electron-phonon system. We note that as a distribution-averaged quantity, $\left<\tau_{r}\right>$ has a similar definition to $\tau_{qp}$ calculated above using Eq.~\ref{eqn:Nqp_solved}, however, $\left<\tau^*_{r}\right>/\left<\tau_{r}\right>$ is not expected to equal $F^*_\phi/F_\phi$ since in Eq.~\ref{eqn:taurKaplanE} there is no signal power, only quasiparticles and phonons in the superconductor are considered, and the system is in equilibrium.

\subsection{Results\label{sec:results}}

\begin{figure}[t]
\begin{center}
\includegraphics[width=7.5cm]{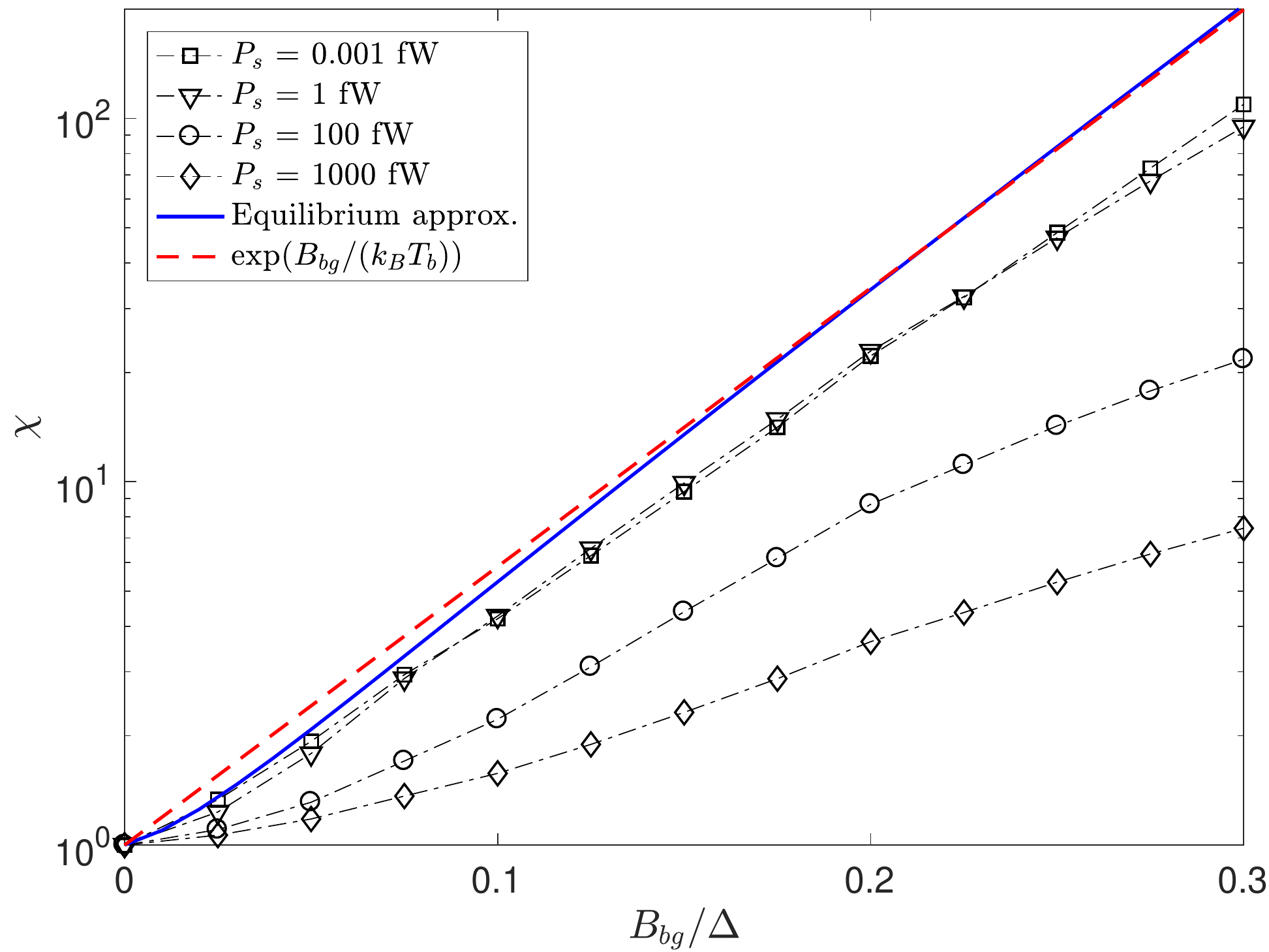}
\caption{\label{fig:chi}~Enhancement of the average quasiparticle recombination time as a function of signal power and phonon bandgap width, $B_{bg}$. For the non-equilibrium results (symbols with dot-dashed lines), the bandgap extends from 2$\Delta$ to $B_{bg}$, which numerically approximates the equilibrium calculation (solid line) given by Eq.~\ref{eqn:taurKaplanE}. The bandgap rejection level, $r_{bg}$, is 10$^4$.} 
\end{center}
\end{figure}

Figure~\ref{fig:chi} shows that more than an order of magnitude increase in the quasiparticle lifetime is possible with a phonon bandgap that can be practically achieved with a PnC. The non-equilibrium calculation of the electron-phonon model in Fig.~\ref{fig:KID-model} shows a decrease in $\chi$ with increasing photon pair-breaking power, an effect that is largely attributed to an increase in the phonon flow to the bath at all energies, and a redistribution of the quasiparticle energies through electron-phonon scattering. 

When $\Gamma_s \ll \Gamma_{pb}$ (small-signal limit), the equilibrium approximation to the quasiparticle recombination time agrees remarkably well with a calculation using the full non-equilibrium solutions of the kinetic equations. The enhancement in $\tau_{qp}$ calculated from Eq.~\ref{eqn:taurKaplanaverage} closely follows an exponential function with $B_{bg}$, i.e. $\chi_r \simeq \exp(B_{bg}/k_BT_b)$, an effect that arises from the Fermi-Dirac and Bose-Einstein distributions for the quasiparticle and phonon occupations, respectively. For the same reason, the effect of a phonon bandgap depends strongly on bath temperature and the density of thermally generated quasiparticles. It is important to note that the estimate of the quasiparticle enhancement from Eq.~\ref{eqn:taurKaplanaverage} will not reach an asymptote as a function of bandgap bandwidth, because the phonon bandgap as introduced in Eq.~\ref{eqn:taurKaplanE} prevents electron-phonon scattering from taking place inside the superconductor at the bandgap energies.

\begin{figure}[htbp]
\begin{center}
\includegraphics[width=8.5cm]{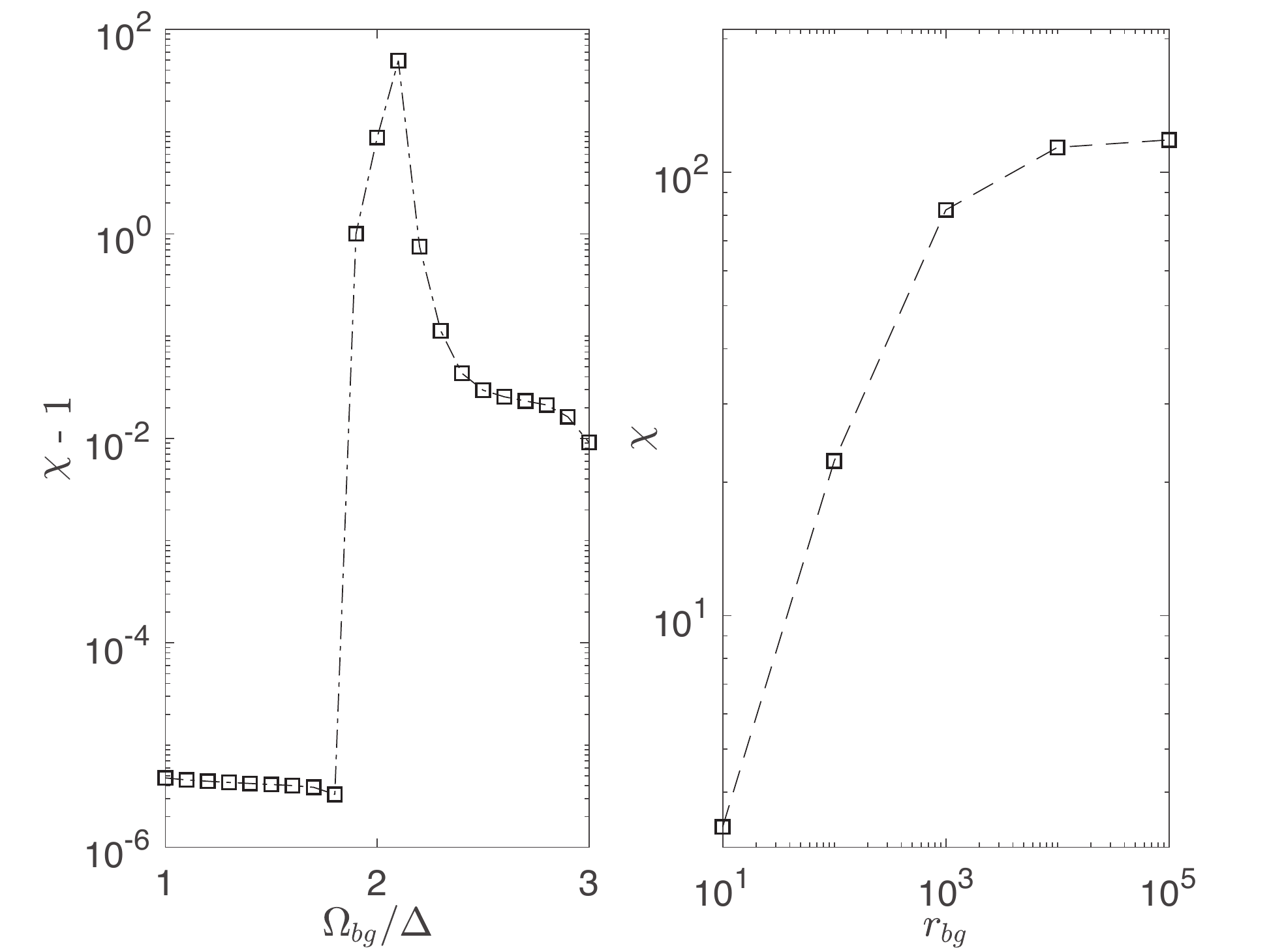}
\caption{(Left) Quasiparticle recombination time enhancement in the small-signal limit and as function of the bandgap position. The sharp increase in $\chi$ is due to the peak in the phonon density of states at $\sim$2$\Delta$ due to quasiparticle recombination. The signal energy is 5$\Delta$, and power is 1 aW/$\mu$m$^3$. The bandgap width is 0.3$\Delta$, and rejection level is 10$^4$. (Right) Recombination time enhancement as a function of bandgap rejection level.}
\label{fig:chi_BGcenter}
\end{center}
\end{figure}

In Fig.~\ref{fig:chi_BGcenter}, the enhancement in the quasiparticle lifetime is calculated as a function of the bandgap center $\Omega_{bg}$, with the bandgap bandwidth $B_{bg}$ fixed to 0.3$\Delta$. The sharp increase in the recombination time at 2$\Delta$ reflects the peak in the quasiparticle density of states near $\Delta$, and peak phonon power flow near 2$\Delta$ due to recombination. Figure~\ref{fig:chi_BGcenter} also shows that capturing the peak of the phonon distribution due recombination (including the exponential decrease thereafter) is essential as $\chi$ reaches a maximum when $\Omega_{bg}$ is at $\sim$2.1$\Delta$. 

It is important to consider how $\chi$ scales with $r_{bg}$. An ideal PnC infinite in extent can be designed to form a complete phonon bandgap, in which case $r_{bg}$ is infinite. Practical constraints in a superconducting circuit will pose a limitation on the PnC size. A truncation in the size of the PnC leads to a bandgap, the rejection level of which depends on the crystal dimensions. The rejection level is typically $10^3$ with 3 units cells, and greater than $10^6$ with 6 or more unit cells.~\cite{rostem-aperiodic,30percent} Figure~\ref{fig:chi_BGcenter} shows that $\chi$ reaches an asymptote with $r_{bg}$, reflecting the finite number of $\sim\,$2$\Delta$ phonons in the superconductor that are prevented from flowing to the thermal bath due to the PnC bandgap.

\section{Discussion}

To experimentally address the quasiparticle lifetime enhancement due a PnC bandgap, we envision a superconducting microwave resonator realized as a lumped-element or a distributed transmission line circuit. We consider the resonator realized from a lumped capacitor and inductor, which is compact and desirable for nano-scale fabrication.

The PnC is patterned around the inductor, which is the only element sensitive to the quasiparticle dynamics. The inductor line width is typically 1-2 $\mu$m with a total required membrane size of 10 $\mu$m, which is smaller than the thermal phonon mean-free-path. The quasiparticle recombination time can be probed by either measuring the generation-recombination noise or illuminating the inductor with pair-breaking light~\cite{deVisser-PRL2011}. The capacitive element of the resonator can be realized with a high-$T_c$ ($>\,$9 K) superconductor (e.g. Nb or NbTiN), which confines the quasiparticles to the inductor metal with a much lower $T_c$. 

The PnC can be realized in a thin dielectric membrane ($\sim\,$100 nm). Current fabrication technologies, such electron-beam lithography and focused ion-beam milling can be used to pattern the membrane into a PnC. The minimum unit cell is limited to $\sim\,$70 nm. Only a few unit cells are needed to realize a greater than $10^4$ suppression of the phonon flux to the bath at the bandgap energies of the PnC. Figure~\ref{fig:KID-model} illustrates a typical configuration. The minimum unit cell limits the center of the bandgap to frequencies $<\,$30 GHz~\cite{anufriev-theory,Tuomas}, which requires a superconducting metal with a $T_c <\,$400 mK (e.g. Ti, Hf, AlMn), that is operated at around 50 mK to suppress thermally generated quasiparticles. 

The sensitivity of the quasiparticle recombination time to a phonon bandgap presents an interesting phenomenon for measuring the spectral response of PnCs using a superconducting microwave resonator. Although there are well established methods for measuring the elastic properties of PnCs, such as Brillouin scattering, coupling a superconducting resonator to a PnC could provide additional insight into the phonon coupling between a superconducting film and an elastic dielectric membrane at sub-Kelvin temperatures. The electro-mechanical system described here would be particularly sensitive to phonon dimensionality, and phonon density of states in each film through the electron-phonon interaction function $\alpha^2(\Omega)\rho_\phi(\Omega)$. Conversely, understanding the electron-phonon coupling in superconducting microwave resonators could be valuable for understanding and pushing the performance limit of current ultra-sensitive microwave kinetic-inductance detectors~\cite{deVisser-Ncomms}.

\section{Conclusions}

We have shown that a phonon bandgap restricting the escape rate of recombination phonons from a superconducting film has a strong effect on the quasiparticle dynamics. The quasiparticle lifetime, $\tau_{qp}$, is increased more than an order of magnitude for realistic bandgap properties that can be achieved with phononic crystals. When the pair-breaking photon rate is low, the enhancement in the quasiparticle lifetime calculated using non-equilibrium particle distributions is in good quantitative agreement with an equilibrium formulation. The enhancement in $\tau_{qp}$ should have a significant impact in superconducting devices where quasiparticle fluctuation plays an important role in limiting the sensitivity and responsivity to pair-breaking photons. In addition, $\tau_{qp}$ is sensitive to the position of the phonon bandgap. This response may provide a useful probe for phonon coupling in low-dimensional thin-films. By symmetry of the phonon transport across a PnC, an isolated superconductor will be less susceptible to pair-breaking phonon-mediated quasiparticle poisoning. 

\begin{acknowledgments}
We gratefully acknowledge financial support from the NASA Astrophysics Research and Analysis (NNX17AH83G) grant and NASA Science Investment Fund. P. J. de Visser is financially supported by the Netherlands Organisation for Scientific Research NWO (Veni Grant 639.041.750). We thank Ilari Maasilta, Jochem Baselmans, Omid Noroozian and Tuomas Puurtinen for useful discussions.
\end{acknowledgments}

\newpage
\section*{Appendix}

The kinetic equations describing the rate of change of quasiparticle and phonon distributions in the superconductor are
\begin{eqnarray}\label{fdot}
\dot{f}(E) = I_{qp} &-& \frac{1}{\tau_0(k_BT_c)^3} \int_0^\infty d\Omega\,\Omega^2 \rho(E+\Omega)   \label{eqn:f_E_rate_apndx} \\
	&\times& L_{-,+}(E,\Omega)[f(E)(1-f(E+\Omega))n(\Omega) \nonumber \\
	&-& (1-f(E))f(E+\Omega)(n(\Omega)+1)] \nonumber\\
	&-& \frac{1}{\tau_0(k_BT_c)^3} \int_0^{E-\Delta} d\Omega\,\Omega^2 \rho(E-\Omega) L_{-,-}(E,\Omega) \nonumber\\
	&\times&  [f(E)(1-f(E-\Omega))(n(\Omega)+1) \nonumber \\
	&-& (1-f(E))f(E-\Omega)n(\Omega)] \nonumber\\
	&-& \frac{1}{\tau_0(k_BT_c)^3} \int_{E+\Delta}^\infty d\Omega\,\Omega^2 \rho(\Omega-E) L_{+,-}(E,\Omega) \nonumber\\
	&\times& [f(E)f(\Omega-E)(n(\Omega)+1) \nonumber \\
	&-& (1-f(E))(1-f(\Omega-E))n(\Omega)] \nonumber
\end{eqnarray}
and
\begin{eqnarray}\label{ndot}
\dot{n}(\Omega) = &-&  \frac{2}{\pi\tau^\phi_0\Delta(0)}  \int_\Delta^\infty dE\rho(E) \rho(E+\Omega) \\
	&\times&  L_{-,+}(E,\Omega)[f(E)(1-f(E+\Omega))n(\Omega) \nonumber \\ 
	&-& (1-f(E))f(E+\Omega)(n(\Omega)+1)] \nonumber\\
	&-& \frac{1}{\pi\tau^\phi_0\Delta(0)}  \int_\Delta^{\Omega-\Delta} dE \rho(E)\,\rho(\Omega-E) L_{+,-}(E,\Omega) \nonumber\\
	&\times&  [(1-f(E))(1-f(\Omega-E))n(\Omega) \nonumber \\ 
	&-& f(E)f(\Omega-E)(n(\Omega)+1)] - \frac{n(\Omega)-m(\Omega)}{\tau_{K}} \nonumber
\end{eqnarray}
respectively~\cite{goldie}. $L_{\pm,\mp}(E,\Omega)=(1\pm\Delta^2/[E(\Omega\mp E))]$. The quasiparticle density of states is $\rho_{qp}(E) = \Re[(E+i\xi)/\sqrt{(E+i\xi)^2-\Delta^2}]$, which accounts for gap smearing through $\xi$. We set $\xi$ to $10^{-5}\Delta$ to ensure numerical convergence of the iterative solver~\cite{goldie}. The rate equation for the phonons in the membrane is given by
\begin{eqnarray}\label{mdot}
\dot{m}(\Omega) &=& -\frac{m(\Omega)-n(\Omega)}{\tau_{K}} - \frac{m(\Omega)-g(\Omega,T_b)}{\tau_{\phi}(\Omega)}.
\end{eqnarray}

In Eq.~\ref{eqn:f_E_rate_apndx}, $I_{qp}$ is the rate of quasiparticle generation from pair-breaking signal photons~\cite{goldie},
\begin{eqnarray}
I_{qp}(E,\nu_s) &=& 2\mathcal{B} \{L_{+,-} (E,\hbar\omega) \rho(E-\hbar\omega) \nonumber \\
			&\times& [f(E-\hbar\omega) - f(E)] \nonumber \\ 
			&-& L_{+,+}(E,\hbar\omega) \rho(E+\hbar\omega)[f(E)-F(E-\hbar\omega)] \nonumber \\
			&+& L_{-,-}(\hbar\omega,E) \rho(\hbar\omega-E) \nonumber \\ 
			&\times& [1-f(E)-f(\hbar\omega-E)] \}, \nonumber 
\end{eqnarray}
where $\mathcal{B}$ is a constant that is determined by satisfying the energy conservation equation in steady-state,
\begin{eqnarray}\label{eqn:Ps-qp}
P_{s} = 4N(0)\mathcal{B}\int_{\Delta}^\infty K_{qp} E\,\rho_{qp}(E)\,dE,
\end{eqnarray}
where $P_s$ is the pair-breaking photon power (see Fig.~\ref{fig:KID-model}). Similarly, the phonon power that flows from the superconductor to the membrane is given by
\begin{eqnarray}\label{eqn:Pn-m}
P_{qp-n} = \mathcal{H} \int_0^\infty \Omega\rho_K(\Omega) \frac{n(\Omega)-m(\Omega)}{\tau_{K}}d\Omega,
\end{eqnarray}
and the phonon power that flows from the membrane to the bath is given by
\begin{eqnarray}\label{eqn:Pm-bath}
P_{m-b} = \mathcal{G} \int_0^\infty \Omega\rho_\phi(\Omega) \frac{m(\Omega)-g(\Omega,T_b)}{\tilde{\tau}_\phi(\Omega)}d\Omega. 
\end{eqnarray}
$\rho_K(\Omega)$ and $\rho_\phi(\Omega)$ are the density of states for the phonon transport across the Kapitza boundary between the superconductor and the membrane, and the membrane and thermal bath, respectively. For simplicity, we assume $\rho_K=\rho_\phi=\rho_D$, where $\rho_D = 9 N_{ion}\Omega^2/\Omega_D^3$ and $\Omega_D$ is a Debye phonon frequency. The assumption that $\rho_K=\rho_D$ has been used to predict the response of a superconducting film to optical pair-breaking photons~\cite{goldie}. In theory, $\rho_\phi$ can be determined for any given infinitely periodic crystal, however, it is significantly more challenging to compute the phonon relaxation time as a function of frequency, which is beyond the scope of this work. We emphasize that $\chi$ is a normalized parameter, which measures an increase in $\tau_{qp}$ relative to a superconducting system without a phononic bandgap. 

We employ and extend the iterative technique described in Ref~\onlinecite{goldie} to solve the set of non-linear equations~\ref{fdot}--\ref{eqn:Pm-bath} in steady-state. Since the density of states and the associated phonon power transmission factors are difficult to determine accurately for a given system, the constants $\mathcal{B}$, $\mathcal{H}$, and $\mathcal{G}$ are solved simultaneously with the particle occupations $f(E)$, $n(\Omega)$, and $m(\Omega)$ for a self consistent solution that conserves energy. Table~\ref{tbl:summary} summarizes the superconducting materials parameters used to model the electron-phonon system shown in Fig.~\ref{fig:KID-model}. 

\begin{table}[h]
\caption{Intrinsic parameters of the superconducting metal (Al) and membrane described in the text. The bath temperature of the system is 0.1\,$T_c$.}
\begin{center}
\begin{tabular}{|c|l|l|}
\hline
\hline
$T_c$ &\, 1.18 &\, [K] \\
$\Delta$ &\, 180 &\, [$\mu$eV] \\
$\xi$ &\, 1.1$\times$10$^{-3}$ &\, [$\Delta$] \\
$N(0)$ &\, 1.7$\times$10$^6$  &\,  [$\mu$eV$^{-1}\mu$m$^{-3}$] \\
$b$ &\, 10$^{-5}$ &\, [$\Delta^{-2}$] \\
$N_{ion}$ &\, 7.1$\times$10$^{10}$  &\, [$\mu$m$^{-3}$] \\
$\Omega_D$  &\,  37 &\, [meV] \\
$\tau_0$ &\, 438 &\, [ns] \\
$\tau^\phi_0$ &\,  0.26 &\, [ns] \\
$\tau_K$  &\, 0.26 &\, [ns] \\
\hline
\end{tabular}
\end{center}
\label{tbl:summary}
\end{table}%

\bibliography{ROSTEM_Phononic_Isolated_KID_PRB.bbl}

\begin{thebibliography}{34}%
\makeatletter
\providecommand \@ifxundefined [1]{%
 \@ifx{#1\undefined}
}%
\providecommand \@ifnum [1]{%
 \ifnum #1\expandafter \@firstoftwo
 \else \expandafter \@secondoftwo
 \fi
}%
\providecommand \@ifx [1]{%
 \ifx #1\expandafter \@firstoftwo
 \else \expandafter \@secondoftwo
 \fi
}%
\providecommand \natexlab [1]{#1}%
\providecommand \enquote  [1]{``#1''}%
\providecommand \bibnamefont  [1]{#1}%
\providecommand \bibfnamefont [1]{#1}%
\providecommand \citenamefont [1]{#1}%
\providecommand \href@noop [0]{\@secondoftwo}%
\providecommand \href [0]{\begingroup \@sanitize@url \@href}%
\providecommand \@href[1]{\@@startlink{#1}\@@href}%
\providecommand \@@href[1]{\endgroup#1\@@endlink}%
\providecommand \@sanitize@url [0]{\catcode `\\12\catcode `\$12\catcode
  `\&12\catcode `\#12\catcode `\^12\catcode `\_12\catcode `\%12\relax}%
\providecommand \@@startlink[1]{}%
\providecommand \@@endlink[0]{}%
\providecommand \url  [0]{\begingroup\@sanitize@url \@url }%
\providecommand \@url [1]{\endgroup\@href {#1}{\urlprefix }}%
\providecommand \urlprefix  [0]{URL }%
\providecommand \Eprint [0]{\href }%
\providecommand \doibase [0]{http://dx.doi.org/}%
\providecommand \selectlanguage [0]{\@gobble}%
\providecommand \bibinfo  [0]{\@secondoftwo}%
\providecommand \bibfield  [0]{\@secondoftwo}%
\providecommand \translation [1]{[#1]}%
\providecommand \BibitemOpen [0]{}%
\providecommand \bibitemStop [0]{}%
\providecommand \bibitemNoStop [0]{.\EOS\space}%
\providecommand \EOS [0]{\spacefactor3000\relax}%
\providecommand \BibitemShut  [1]{\csname bibitem#1\endcsname}%
\let\auto@bib@innerbib\@empty
\bibitem [{\citenamefont {Alegre}\ \emph {et~al.}(2011)\citenamefont {Alegre},
  \citenamefont {Safavi-Naeini}, \citenamefont {Winger},\ and\ \citenamefont
  {Painter}}]{alegre}%
  \BibitemOpen
  \bibfield  {author} {\bibinfo {author} {\bibfnamefont {T.~P.~M.}\
  \bibnamefont {Alegre}}, \bibinfo {author} {\bibfnamefont {A.}~\bibnamefont
  {Safavi-Naeini}}, \bibinfo {author} {\bibfnamefont {M.}~\bibnamefont
  {Winger}}, \ and\ \bibinfo {author} {\bibfnamefont {O.}~\bibnamefont
  {Painter}},\ }\href@noop {} {\bibfield  {journal} {\bibinfo  {journal} {Opt.
  Express}\ }\textbf {\bibinfo {volume} {19}},\ \bibinfo {pages} {5658}
  (\bibinfo {year} {2011})}\BibitemShut {NoStop}%
\bibitem [{\citenamefont {Schmidt}\ \emph {et~al.}(2012)\citenamefont
  {Schmidt}, \citenamefont {Ludwig},\ and\ \citenamefont
  {Marquardt}}]{QIprocessing}%
  \BibitemOpen
  \bibfield  {author} {\bibinfo {author} {\bibfnamefont {M.}~\bibnamefont
  {Schmidt}}, \bibinfo {author} {\bibfnamefont {M.}~\bibnamefont {Ludwig}}, \
  and\ \bibinfo {author} {\bibfnamefont {F.}~\bibnamefont {Marquardt}},\ }\href
  {http://stacks.iop.org/1367-2630/14/i=12/a=125005} {\bibfield  {journal}
  {\bibinfo  {journal} {New J. Phys.}\ }\textbf {\bibinfo {volume} {14}},\
  \bibinfo {pages} {125005} (\bibinfo {year} {2012})}\BibitemShut {NoStop}%
\bibitem [{\citenamefont {Chan}\ \emph {et~al.}(2012)\citenamefont {Chan},
  \citenamefont {Safavi-Naeini}, \citenamefont {Hill}, \citenamefont
  {Meenehan},\ and\ \citenamefont {Painter}}]{Chan-AcousticShield}%
  \BibitemOpen
  \bibfield  {author} {\bibinfo {author} {\bibfnamefont {J.}~\bibnamefont
  {Chan}}, \bibinfo {author} {\bibfnamefont {A.~H.}\ \bibnamefont
  {Safavi-Naeini}}, \bibinfo {author} {\bibfnamefont {J.~T.}\ \bibnamefont
  {Hill}}, \bibinfo {author} {\bibfnamefont {S.}~\bibnamefont {Meenehan}}, \
  and\ \bibinfo {author} {\bibfnamefont {O.}~\bibnamefont {Painter}},\
  }\href@noop {} {\bibfield  {journal} {\bibinfo  {journal} {Appl. Phys.
  Lett.}\ }\textbf {\bibinfo {volume} {101}},\ \bibinfo {pages} {081115}
  (\bibinfo {year} {2012})}\BibitemShut {NoStop}%
\bibitem [{\citenamefont {Sukhovich}\ \emph {et~al.}(2013)\citenamefont
  {Sukhovich}, \citenamefont {Page}, \citenamefont {Vasseur}, \citenamefont
  {Robillard}, \citenamefont {Swinteck},\ and\ \citenamefont
  {Deymier}}]{30percent}%
  \BibitemOpen
  \bibfield  {author} {\bibinfo {author} {\bibfnamefont {A.}~\bibnamefont
  {Sukhovich}}, \bibinfo {author} {\bibfnamefont {J.~H.}\ \bibnamefont {Page}},
  \bibinfo {author} {\bibfnamefont {J.~O.}\ \bibnamefont {Vasseur}}, \bibinfo
  {author} {\bibfnamefont {J.~F.}\ \bibnamefont {Robillard}}, \bibinfo {author}
  {\bibfnamefont {N.}~\bibnamefont {Swinteck}}, \ and\ \bibinfo {author}
  {\bibfnamefont {P.~A.}\ \bibnamefont {Deymier}},\ }in\ \href@noop {} {\emph
  {\bibinfo {booktitle} {Acoustic Metamaterials and Phononic Crystals}}},\
  \bibinfo {series} {Springer Series in Solid-State Sciences}, Vol.\ \bibinfo
  {volume} {173}\ (\bibinfo  {publisher} {Springer Berlin Heidelberg},\
  \bibinfo {year} {2013})\ pp.\ \bibinfo {pages} {95--157}\BibitemShut
  {NoStop}%
\bibitem [{\citenamefont {Safavi-Naeini}\ and\ \citenamefont
  {Painter}(2010)}]{amir:2010}%
  \BibitemOpen
  \bibfield  {author} {\bibinfo {author} {\bibfnamefont {A.~H.}\ \bibnamefont
  {Safavi-Naeini}}\ and\ \bibinfo {author} {\bibfnamefont {O.}~\bibnamefont
  {Painter}},\ }\href {\doibase 10.1364/OE.18.014926} {\bibfield  {journal}
  {\bibinfo  {journal} {Opt. Express}\ }\textbf {\bibinfo {volume} {18}},\
  \bibinfo {pages} {14926} (\bibinfo {year} {2010})}\BibitemShut {NoStop}%
\bibitem [{\citenamefont {Wang}\ and\ \citenamefont
  {Li}(2007)}]{phononic:transistor}%
  \BibitemOpen
  \bibfield  {author} {\bibinfo {author} {\bibfnamefont {L.}~\bibnamefont
  {Wang}}\ and\ \bibinfo {author} {\bibfnamefont {B.}~\bibnamefont {Li}},\
  }\href@noop {} {\bibfield  {journal} {\bibinfo  {journal} {Phys. Rev. Lett.}\
  }\textbf {\bibinfo {volume} {99}},\ \bibinfo {pages} {177208} (\bibinfo
  {year} {2007})}\BibitemShut {NoStop}%
\bibitem [{\citenamefont {Yu}\ \emph {et~al.}(2014)\citenamefont {Yu},
  \citenamefont {Cicak}, \citenamefont {Kampel}, \citenamefont {Tsaturyan},
  \citenamefont {Purdy}, \citenamefont {Simmonds},\ and\ \citenamefont
  {Regal}}]{Yu:phononicShield}%
  \BibitemOpen
  \bibfield  {author} {\bibinfo {author} {\bibfnamefont {P.-L.}\ \bibnamefont
  {Yu}}, \bibinfo {author} {\bibfnamefont {K.}~\bibnamefont {Cicak}}, \bibinfo
  {author} {\bibfnamefont {N.~S.}\ \bibnamefont {Kampel}}, \bibinfo {author}
  {\bibfnamefont {Y.}~\bibnamefont {Tsaturyan}}, \bibinfo {author}
  {\bibfnamefont {T.~P.}\ \bibnamefont {Purdy}}, \bibinfo {author}
  {\bibfnamefont {R.~W.}\ \bibnamefont {Simmonds}}, \ and\ \bibinfo {author}
  {\bibfnamefont {C.~A.}\ \bibnamefont {Regal}},\ }\href@noop {} {\bibfield
  {journal} {\bibinfo  {journal} {Appl. Phys. Lett.}\ }\textbf {\bibinfo
  {volume} {104}} (\bibinfo {year} {2014})}\BibitemShut {NoStop}%
\bibitem [{\citenamefont {Zen}\ \emph {et~al.}(2014)\citenamefont {Zen},
  \citenamefont {Puurtinen}, \citenamefont {Isotalo}, \citenamefont
  {Chaudhuri},\ and\ \citenamefont {Maasilta}}]{zen}%
  \BibitemOpen
  \bibfield  {author} {\bibinfo {author} {\bibfnamefont {N.}~\bibnamefont
  {Zen}}, \bibinfo {author} {\bibfnamefont {T.~A.}\ \bibnamefont {Puurtinen}},
  \bibinfo {author} {\bibfnamefont {T.~J.}\ \bibnamefont {Isotalo}}, \bibinfo
  {author} {\bibfnamefont {S.}~\bibnamefont {Chaudhuri}}, \ and\ \bibinfo
  {author} {\bibfnamefont {I.~J.}\ \bibnamefont {Maasilta}},\ }\href@noop {}
  {\bibfield  {journal} {\bibinfo  {journal} {Nat. Commun.}\ }\textbf {\bibinfo
  {volume} {5}} (\bibinfo {year} {2014})}\BibitemShut {NoStop}%
\bibitem [{\citenamefont {Rostem}\ \emph {et~al.}(2016)\citenamefont {Rostem},
  \citenamefont {Chuss}, \citenamefont {Denis},\ and\ \citenamefont
  {Wollack}}]{rostem-aperiodic}%
  \BibitemOpen
  \bibfield  {author} {\bibinfo {author} {\bibfnamefont {K.}~\bibnamefont
  {Rostem}}, \bibinfo {author} {\bibfnamefont {D.~T.}\ \bibnamefont {Chuss}},
  \bibinfo {author} {\bibfnamefont {K.~L.}\ \bibnamefont {Denis}}, \ and\
  \bibinfo {author} {\bibfnamefont {E.~J.}\ \bibnamefont {Wollack}},\
  }\href@noop {} {\bibfield  {journal} {\bibinfo  {journal} {J. Phys. D: Appl.
  Phys.}\ }\textbf {\bibinfo {volume} {49}},\ \bibinfo {pages} {255301}
  (\bibinfo {year} {2016})}\BibitemShut {NoStop}%
\bibitem [{\citenamefont {Anufriev}\ \emph {et~al.}(2016)\citenamefont
  {Anufriev}, \citenamefont {Maire},\ and\ \citenamefont
  {Nomura}}]{anufriev-meas}%
  \BibitemOpen
  \bibfield  {author} {\bibinfo {author} {\bibfnamefont {R.}~\bibnamefont
  {Anufriev}}, \bibinfo {author} {\bibfnamefont {J.}~\bibnamefont {Maire}}, \
  and\ \bibinfo {author} {\bibfnamefont {M.}~\bibnamefont {Nomura}},\ }\href
  {\doibase 10.1103/PhysRevB.93.045411} {\bibfield  {journal} {\bibinfo
  {journal} {Phys. Rev. B}\ }\textbf {\bibinfo {volume} {93}},\ \bibinfo
  {pages} {045411} (\bibinfo {year} {2016})}\BibitemShut {NoStop}%
\bibitem [{\citenamefont {Chang}\ and\ \citenamefont
  {Scalapino}(1978)}]{chang_NE}%
  \BibitemOpen
  \bibfield  {author} {\bibinfo {author} {\bibfnamefont {J.-J.}\ \bibnamefont
  {Chang}}\ and\ \bibinfo {author} {\bibfnamefont {D.~J.}\ \bibnamefont
  {Scalapino}},\ }\href {\doibase 10.1007/BF00116228} {\bibfield  {journal}
  {\bibinfo  {journal} {J. Low Tem. Phys.}\ }\textbf {\bibinfo {volume} {31}},\
  \bibinfo {pages} {1} (\bibinfo {year} {1978})}\BibitemShut {NoStop}%
\bibitem [{\citenamefont {Wilson}\ and\ \citenamefont {Prober}(2004)}]{wilson}%
  \BibitemOpen
  \bibfield  {author} {\bibinfo {author} {\bibfnamefont {C.~M.}\ \bibnamefont
  {Wilson}}\ and\ \bibinfo {author} {\bibfnamefont {D.~E.}\ \bibnamefont
  {Prober}},\ }\href@noop {} {\bibfield  {journal} {\bibinfo  {journal} {Phys.
  Rev. B}\ }\textbf {\bibinfo {volume} {69}},\ \bibinfo {pages} {094524}
  (\bibinfo {year} {2004})}\BibitemShut {NoStop}%
\bibitem [{\citenamefont {Day}\ \emph {et~al.}(2003)\citenamefont {Day},
  \citenamefont {LeDuc}, \citenamefont {Mazin}, \citenamefont {Vayonakis},\
  and\ \citenamefont {Zmuidzinas}}]{day}%
  \BibitemOpen
  \bibfield  {author} {\bibinfo {author} {\bibfnamefont {P.~K.}\ \bibnamefont
  {Day}}, \bibinfo {author} {\bibfnamefont {H.~G.}\ \bibnamefont {LeDuc}},
  \bibinfo {author} {\bibfnamefont {B.~A.}\ \bibnamefont {Mazin}}, \bibinfo
  {author} {\bibfnamefont {A.}~\bibnamefont {Vayonakis}}, \ and\ \bibinfo
  {author} {\bibfnamefont {J.}~\bibnamefont {Zmuidzinas}},\ }\href@noop {}
  {\bibfield  {journal} {\bibinfo  {journal} {Nature}\ }\textbf {\bibinfo
  {volume} {425}},\ \bibinfo {pages} {817} (\bibinfo {year}
  {2003})}\BibitemShut {NoStop}%
\bibitem [{\citenamefont {Zmuidzinas}(2012)}]{jonas}%
  \BibitemOpen
  \bibfield  {author} {\bibinfo {author} {\bibfnamefont {J.}~\bibnamefont
  {Zmuidzinas}},\ }\href {\doibase 10.1146/annurev-conmatphys-020911-125022}
  {\bibfield  {journal} {\bibinfo  {journal} {Annu. Rev. Condens. Matter
  Phys.}\ }\textbf {\bibinfo {volume} {3}},\ \bibinfo {pages} {169} (\bibinfo
  {year} {2012})}\BibitemShut {NoStop}%
\bibitem [{\citenamefont {de~Visser}\ \emph {et~al.}(2014)\citenamefont
  {de~Visser}, \citenamefont {Baselmans}, \citenamefont {Bueno}, \citenamefont
  {Llombart},\ and\ \citenamefont {Klapwijk}}]{deVisser-Ncomms}%
  \BibitemOpen
  \bibfield  {author} {\bibinfo {author} {\bibfnamefont {P.~J.}\ \bibnamefont
  {de~Visser}}, \bibinfo {author} {\bibfnamefont {J.~J.~A.}\ \bibnamefont
  {Baselmans}}, \bibinfo {author} {\bibfnamefont {J.}~\bibnamefont {Bueno}},
  \bibinfo {author} {\bibfnamefont {N.}~\bibnamefont {Llombart}}, \ and\
  \bibinfo {author} {\bibfnamefont {T.~M.}\ \bibnamefont {Klapwijk}},\
  }\href@noop {} {\bibfield  {journal} {\bibinfo  {journal} {Nat. Commun.}\
  }\textbf {\bibinfo {volume} {5}},\ \bibinfo {pages} {3130} (\bibinfo {year}
  {2014})}\BibitemShut {NoStop}%
\bibitem [{\citenamefont {Shaw}\ \emph {et~al.}(2009)\citenamefont {Shaw},
  \citenamefont {Bueno}, \citenamefont {Day}, \citenamefont {Bradford},\ and\
  \citenamefont {Echternach}}]{QCD-theory}%
  \BibitemOpen
  \bibfield  {author} {\bibinfo {author} {\bibfnamefont {M.~D.}\ \bibnamefont
  {Shaw}}, \bibinfo {author} {\bibfnamefont {J.}~\bibnamefont {Bueno}},
  \bibinfo {author} {\bibfnamefont {P.}~\bibnamefont {Day}}, \bibinfo {author}
  {\bibfnamefont {C.~M.}\ \bibnamefont {Bradford}}, \ and\ \bibinfo {author}
  {\bibfnamefont {P.~M.}\ \bibnamefont {Echternach}},\ }\href@noop {}
  {\bibfield  {journal} {\bibinfo  {journal} {Phys. Rev. B}\ }\textbf {\bibinfo
  {volume} {79}},\ \bibinfo {pages} {144511} (\bibinfo {year}
  {2009})}\BibitemShut {NoStop}%
\bibitem [{\citenamefont {Echternach}\ \emph {et~al.}(2018)\citenamefont
  {Echternach}, \citenamefont {Pepper}, \citenamefont {Reck},\ and\
  \citenamefont {Bradford}}]{QCD-2018}%
  \BibitemOpen
  \bibfield  {author} {\bibinfo {author} {\bibfnamefont {P.~M.}\ \bibnamefont
  {Echternach}}, \bibinfo {author} {\bibfnamefont {B.~J.}\ \bibnamefont
  {Pepper}}, \bibinfo {author} {\bibfnamefont {T.}~\bibnamefont {Reck}}, \ and\
  \bibinfo {author} {\bibfnamefont {C.~M.}\ \bibnamefont {Bradford}},\
  }\href@noop {} {\bibfield  {journal} {\bibinfo  {journal} {Nat. Astron.}\
  }\textbf {\bibinfo {volume} {2}},\ \bibinfo {pages} {90} (\bibinfo {year}
  {2018})}\BibitemShut {NoStop}%
\bibitem [{\citenamefont {Patel}\ \emph {et~al.}(2017)\citenamefont {Patel},
  \citenamefont {Pechenezhskiy}, \citenamefont {Plourde}, \citenamefont
  {Vavilov},\ and\ \citenamefont {McDermott}}]{QUBIT-poisening}%
  \BibitemOpen
  \bibfield  {author} {\bibinfo {author} {\bibfnamefont {U.}~\bibnamefont
  {Patel}}, \bibinfo {author} {\bibfnamefont {I.~V.}\ \bibnamefont
  {Pechenezhskiy}}, \bibinfo {author} {\bibfnamefont {B.~L.~T.}\ \bibnamefont
  {Plourde}}, \bibinfo {author} {\bibfnamefont {M.~G.}\ \bibnamefont
  {Vavilov}}, \ and\ \bibinfo {author} {\bibfnamefont {R.}~\bibnamefont
  {McDermott}},\ }\href {\doibase 10.1103/PhysRevB.96.220501} {\bibfield
  {journal} {\bibinfo  {journal} {Phys. Rev. B}\ }\textbf {\bibinfo {volume}
  {96}},\ \bibinfo {pages} {220501} (\bibinfo {year} {2017})}\BibitemShut
  {NoStop}%
\bibitem [{\citenamefont {Vercruyssen}\ \emph {et~al.}(2011)\citenamefont
  {Vercruyssen}, \citenamefont {Barends}, \citenamefont {Klapwijk},
  \citenamefont {Muhonen}, \citenamefont {Meschke},\ and\ \citenamefont
  {Pekola}}]{vercruyssen}%
  \BibitemOpen
  \bibfield  {author} {\bibinfo {author} {\bibfnamefont {N.}~\bibnamefont
  {Vercruyssen}}, \bibinfo {author} {\bibfnamefont {R.}~\bibnamefont
  {Barends}}, \bibinfo {author} {\bibfnamefont {T.~M.}\ \bibnamefont
  {Klapwijk}}, \bibinfo {author} {\bibfnamefont {J.~T.}\ \bibnamefont
  {Muhonen}}, \bibinfo {author} {\bibfnamefont {M.}~\bibnamefont {Meschke}}, \
  and\ \bibinfo {author} {\bibfnamefont {J.~P.}\ \bibnamefont {Pekola}},\
  }\href {\doibase 10.1063/1.3624463} {\bibfield  {journal} {\bibinfo
  {journal} {Appl. Phys. Lett.}\ }\textbf {\bibinfo {volume} {99}},\ \bibinfo
  {pages} {062509} (\bibinfo {year} {2011})}\BibitemShut {NoStop}%
\bibitem [{\citenamefont {Fyhrie}\ \emph {et~al.}(2016)\citenamefont {Fyhrie},
  \citenamefont {McKenney}, \citenamefont {Glenn}, \citenamefont {LeDuc},
  \citenamefont {Gao}, \citenamefont {Day},\ and\ \citenamefont
  {Zmuidzinas}}]{fyhrie}%
  \BibitemOpen
  \bibfield  {author} {\bibinfo {author} {\bibfnamefont {A.}~\bibnamefont
  {Fyhrie}}, \bibinfo {author} {\bibfnamefont {C.}~\bibnamefont {McKenney}},
  \bibinfo {author} {\bibfnamefont {J.}~\bibnamefont {Glenn}}, \bibinfo
  {author} {\bibfnamefont {H.~G.}\ \bibnamefont {LeDuc}}, \bibinfo {author}
  {\bibfnamefont {J.}~\bibnamefont {Gao}}, \bibinfo {author} {\bibfnamefont
  {P.}~\bibnamefont {Day}}, \ and\ \bibinfo {author} {\bibfnamefont
  {J.}~\bibnamefont {Zmuidzinas}},\ }\href@noop {} {\bibfield  {journal}
  {\bibinfo  {journal} {Proc. SPIE}\ }\textbf {\bibinfo {volume} {9914}},\
  \bibinfo {pages} {99142B} (\bibinfo {year} {2016})}\BibitemShut {NoStop}%
\bibitem [{\citenamefont {Klitsner}\ \emph {et~al.}(1988)\citenamefont
  {Klitsner}, \citenamefont {VanCleve}, \citenamefont {Fischer},\ and\
  \citenamefont {Pohl}}]{klitsner}%
  \BibitemOpen
  \bibfield  {author} {\bibinfo {author} {\bibfnamefont {T.}~\bibnamefont
  {Klitsner}}, \bibinfo {author} {\bibfnamefont {J.~E.}\ \bibnamefont
  {VanCleve}}, \bibinfo {author} {\bibfnamefont {H.~E.}\ \bibnamefont
  {Fischer}}, \ and\ \bibinfo {author} {\bibfnamefont {R.~O.}\ \bibnamefont
  {Pohl}},\ }\href@noop {} {\bibfield  {journal} {\bibinfo  {journal} {Phys.
  Rev. B}\ }\textbf {\bibinfo {volume} {38}},\ \bibinfo {pages} {7576}
  (\bibinfo {year} {1988})}\BibitemShut {NoStop}%
\bibitem [{\citenamefont {Holmes}\ \emph {et~al.}(1998)\citenamefont {Holmes},
  \citenamefont {Gildemeister}, \citenamefont {Richards},\ and\ \citenamefont
  {Kotsubo}}]{holmes}%
  \BibitemOpen
  \bibfield  {author} {\bibinfo {author} {\bibfnamefont {W.}~\bibnamefont
  {Holmes}}, \bibinfo {author} {\bibfnamefont {J.~M.}\ \bibnamefont
  {Gildemeister}}, \bibinfo {author} {\bibfnamefont {P.~L.}\ \bibnamefont
  {Richards}}, \ and\ \bibinfo {author} {\bibfnamefont {V.}~\bibnamefont
  {Kotsubo}},\ }\href@noop {} {\bibfield  {journal} {\bibinfo  {journal} {Appl.
  Phys. Lett.}\ }\textbf {\bibinfo {volume} {72}},\ \bibinfo {pages} {2250}
  (\bibinfo {year} {1998})}\BibitemShut {NoStop}%
\bibitem [{\citenamefont {Hoevers}\ \emph {et~al.}(2005)\citenamefont
  {Hoevers}, \citenamefont {Ridder}, \citenamefont {Germeau}, \citenamefont
  {Bruijn}, \citenamefont {de~Korte},\ and\ \citenamefont
  {Wiegerink}}]{hoevers}%
  \BibitemOpen
  \bibfield  {author} {\bibinfo {author} {\bibfnamefont {H.~F.~C.}\
  \bibnamefont {Hoevers}}, \bibinfo {author} {\bibfnamefont {M.~L.}\
  \bibnamefont {Ridder}}, \bibinfo {author} {\bibfnamefont {A.}~\bibnamefont
  {Germeau}}, \bibinfo {author} {\bibfnamefont {M.~P.}\ \bibnamefont {Bruijn}},
  \bibinfo {author} {\bibfnamefont {P.~A.~J.}\ \bibnamefont {de~Korte}}, \ and\
  \bibinfo {author} {\bibfnamefont {R.~J.}\ \bibnamefont {Wiegerink}},\
  }\href@noop {} {\bibfield  {journal} {\bibinfo  {journal} {Appl. Phys.
  Lett.}\ }\textbf {\bibinfo {volume} {86}},\ \bibinfo {pages} {251903}
  (\bibinfo {year} {2005})}\BibitemShut {NoStop}%
\bibitem [{\citenamefont {Lindeman}\ \emph {et~al.}(2014)\citenamefont
  {Lindeman}, \citenamefont {Bonetti}, \citenamefont {Bumble}, \citenamefont
  {Day}, \citenamefont {Eom}, \citenamefont {Holmes},\ and\ \citenamefont
  {Kleinsasser}}]{lindeman}%
  \BibitemOpen
  \bibfield  {author} {\bibinfo {author} {\bibfnamefont {M.~A.}\ \bibnamefont
  {Lindeman}}, \bibinfo {author} {\bibfnamefont {J.~A.}\ \bibnamefont
  {Bonetti}}, \bibinfo {author} {\bibfnamefont {B.}~\bibnamefont {Bumble}},
  \bibinfo {author} {\bibfnamefont {P.~K.}\ \bibnamefont {Day}}, \bibinfo
  {author} {\bibfnamefont {B.~H.}\ \bibnamefont {Eom}}, \bibinfo {author}
  {\bibfnamefont {W.~A.}\ \bibnamefont {Holmes}}, \ and\ \bibinfo {author}
  {\bibfnamefont {A.~W.}\ \bibnamefont {Kleinsasser}},\ }\href {\doibase
  10.1063/1.4884437} {\bibfield  {journal} {\bibinfo  {journal} {J. Appl.
  Phys.}\ }\textbf {\bibinfo {volume} {115}},\ \bibinfo {pages} {234509}
  (\bibinfo {year} {2014})}\BibitemShut {NoStop}%
\bibitem [{\citenamefont {Hou}\ and\ \citenamefont {Assouar}(2008)}]{hou}%
  \BibitemOpen
  \bibfield  {author} {\bibinfo {author} {\bibfnamefont {Z.}~\bibnamefont
  {Hou}}\ and\ \bibinfo {author} {\bibfnamefont {B.~M.}\ \bibnamefont
  {Assouar}},\ }\href@noop {} {\bibfield  {journal} {\bibinfo  {journal} {J.
  Phys. D: Appl. Phys.}\ }\textbf {\bibinfo {volume} {41}},\ \bibinfo {pages}
  {095103} (\bibinfo {year} {2008})}\BibitemShut {NoStop}%
\bibitem [{\citenamefont {Chang}\ and\ \citenamefont
  {Scalapino}(1977)}]{chang_KE}%
  \BibitemOpen
  \bibfield  {author} {\bibinfo {author} {\bibfnamefont {J.-J.}\ \bibnamefont
  {Chang}}\ and\ \bibinfo {author} {\bibfnamefont {D.~J.}\ \bibnamefont
  {Scalapino}},\ }\href {\doibase 10.1103/PhysRevB.15.2651} {\bibfield
  {journal} {\bibinfo  {journal} {Phys. Rev. B}\ }\textbf {\bibinfo {volume}
  {15}},\ \bibinfo {pages} {2651} (\bibinfo {year} {1977})}\BibitemShut
  {NoStop}%
\bibitem [{\citenamefont {Goldie}\ and\ \citenamefont
  {Withington}(2013)}]{goldie}%
  \BibitemOpen
  \bibfield  {author} {\bibinfo {author} {\bibfnamefont {D.~J.}\ \bibnamefont
  {Goldie}}\ and\ \bibinfo {author} {\bibfnamefont {S.}~\bibnamefont
  {Withington}},\ }\href@noop {} {\bibfield  {journal} {\bibinfo  {journal}
  {Supercond. Sci. Tech.}\ }\textbf {\bibinfo {volume} {26}},\ \bibinfo {pages}
  {015004} (\bibinfo {year} {2013})}\BibitemShut {NoStop}%
\bibitem [{\citenamefont {Guruswamy}\ \emph {et~al.}(2014)\citenamefont
  {Guruswamy}, \citenamefont {Goldie},\ and\ \citenamefont
  {Withington}}]{guruswamy_eta}%
  \BibitemOpen
  \bibfield  {author} {\bibinfo {author} {\bibfnamefont {T.}~\bibnamefont
  {Guruswamy}}, \bibinfo {author} {\bibfnamefont {D.~J.}\ \bibnamefont
  {Goldie}}, \ and\ \bibinfo {author} {\bibfnamefont {S.}~\bibnamefont
  {Withington}},\ }\href@noop {} {\bibfield  {journal} {\bibinfo  {journal}
  {Supercond. Sci. Tech.}\ }\textbf {\bibinfo {volume} {27}},\ \bibinfo {pages}
  {055012} (\bibinfo {year} {2014})}\BibitemShut {NoStop}%
\bibitem [{\citenamefont {Kaplan}\ \emph {et~al.}(1976)\citenamefont {Kaplan},
  \citenamefont {Chi}, \citenamefont {Langenberg}, \citenamefont {Chang},
  \citenamefont {Jafarey},\ and\ \citenamefont {Scalapino}}]{kaplan}%
  \BibitemOpen
  \bibfield  {author} {\bibinfo {author} {\bibfnamefont {S.~B.}\ \bibnamefont
  {Kaplan}}, \bibinfo {author} {\bibfnamefont {C.~C.}\ \bibnamefont {Chi}},
  \bibinfo {author} {\bibfnamefont {D.~N.}\ \bibnamefont {Langenberg}},
  \bibinfo {author} {\bibfnamefont {J.~J.}\ \bibnamefont {Chang}}, \bibinfo
  {author} {\bibfnamefont {S.}~\bibnamefont {Jafarey}}, \ and\ \bibinfo
  {author} {\bibfnamefont {D.~J.}\ \bibnamefont {Scalapino}},\ }\href {\doibase
  10.1103/PhysRevB.14.4854} {\bibfield  {journal} {\bibinfo  {journal} {Phys.
  Rev. B}\ }\textbf {\bibinfo {volume} {14}},\ \bibinfo {pages} {4854}
  (\bibinfo {year} {1976})}\BibitemShut {NoStop}%
\bibitem [{\citenamefont {Anufriev}\ and\ \citenamefont
  {Nomura}(2016)}]{anufriev-theory}%
  \BibitemOpen
  \bibfield  {author} {\bibinfo {author} {\bibfnamefont {R.}~\bibnamefont
  {Anufriev}}\ and\ \bibinfo {author} {\bibfnamefont {M.}~\bibnamefont
  {Nomura}},\ }\href {\doibase 10.1103/PhysRevB.93.045410} {\bibfield
  {journal} {\bibinfo  {journal} {Phys. Rev. B}\ }\textbf {\bibinfo {volume}
  {93}},\ \bibinfo {pages} {045410} (\bibinfo {year} {2016})}\BibitemShut
  {NoStop}%
\bibitem [{Tuo()}]{Tuomas}%
  \BibitemOpen
  \href@noop {} {}\bibinfo {note} {T. Puurtinen (private
  communication)}\BibitemShut {NoStop}%
\bibitem [{\citenamefont {Bardeen}\ \emph {et~al.}(1957)\citenamefont
  {Bardeen}, \citenamefont {Cooper},\ and\ \citenamefont {Schrieffer}}]{BCS}%
  \BibitemOpen
  \bibfield  {author} {\bibinfo {author} {\bibfnamefont {J.}~\bibnamefont
  {Bardeen}}, \bibinfo {author} {\bibfnamefont {L.~N.}\ \bibnamefont {Cooper}},
  \ and\ \bibinfo {author} {\bibfnamefont {J.~R.}\ \bibnamefont {Schrieffer}},\
  }\href {\doibase 10.1103/PhysRev.108.1175} {\bibfield  {journal} {\bibinfo
  {journal} {Phys. Rev.}\ }\textbf {\bibinfo {volume} {108}},\ \bibinfo {pages}
  {1175} (\bibinfo {year} {1957})}\BibitemShut {NoStop}%
\bibitem [{\citenamefont {Rothwarf}\ and\ \citenamefont {Taylor}(1967)}]{RT}%
  \BibitemOpen
  \bibfield  {author} {\bibinfo {author} {\bibfnamefont {A.}~\bibnamefont
  {Rothwarf}}\ and\ \bibinfo {author} {\bibfnamefont {B.~N.}\ \bibnamefont
  {Taylor}},\ }\href@noop {} {\bibfield  {journal} {\bibinfo  {journal} {Phys.
  Rev. Lett.}\ }\textbf {\bibinfo {volume} {19}},\ \bibinfo {pages} {27}
  (\bibinfo {year} {1967})}\BibitemShut {NoStop}%
\bibitem [{\citenamefont {de~Visser}\ \emph {et~al.}(2011)\citenamefont
  {de~Visser}, \citenamefont {Baselmans}, \citenamefont {Diener}, \citenamefont
  {Yates}, \citenamefont {Endo},\ and\ \citenamefont
  {Klapwijk}}]{deVisser-PRL2011}%
  \BibitemOpen
  \bibfield  {author} {\bibinfo {author} {\bibfnamefont {P.~J.}\ \bibnamefont
  {de~Visser}}, \bibinfo {author} {\bibfnamefont {J.~J.~A.}\ \bibnamefont
  {Baselmans}}, \bibinfo {author} {\bibfnamefont {P.}~\bibnamefont {Diener}},
  \bibinfo {author} {\bibfnamefont {S.~J.~C.}\ \bibnamefont {Yates}}, \bibinfo
  {author} {\bibfnamefont {A.}~\bibnamefont {Endo}}, \ and\ \bibinfo {author}
  {\bibfnamefont {T.~M.}\ \bibnamefont {Klapwijk}},\ }\href@noop {} {\bibfield
  {journal} {\bibinfo  {journal} {Phys. Rev. Lett.}\ }\textbf {\bibinfo
  {volume} {106}},\ \bibinfo {pages} {167004} (\bibinfo {year}
  {2011})}\BibitemShut {NoStop}%
\end{thebibliography}%

\end{document}